\documentclass[11pt,a4paper]{article}

\usepackage{fullpage}
\usepackage{amsmath}
\usepackage{amssymb}
\usepackage{graphicx}
\usepackage{psfrag}
\usepackage{color}

\newtheorem{definition}{Definition}
\newtheorem{observation}{Observation}
\newtheorem{theorem}{Theorem}
\newtheorem{corollary}{Corollary}
\newtheorem{lemma}{Lemma}
\newenvironment{proof}{\noindent \begin{rm}{\textbf{Proof.} }}{\hspace*{\fill}$\Box$\par\end{rm} \vspace{.3cm}}

\newtheorem{remark}{Remark}

\newcommand{\parent}{\mbox{\sf parent}}
\newcommand{\state}{\mbox{\sf state}}

\newcommand{\next}{\mbox{\sf suc}}
\newcommand{\La}{\mbox{\sf label}}
\newcommand{\dist}{\mbox{\sf d}}

\newcommand{\VarC}{\mbox{\sf VarCycle}}
\newcommand{\DefC}{\mbox{\sf DefCycle}}

\newcommand{\Be}{\mbox{\sf Before}}
\newcommand{\Af}{\mbox{\sf After}}
\newcommand{\End}{\mbox{\sf end}}
\newcommand{\done}{\mbox{\sf done}}

\newcommand{\Ve}{\mbox{\tt Verify}}
\newcommand{\Do}{\mbox{\tt Done}}
\newcommand{\Imp}{\mbox{\tt Improve}}

\newcommand{\Prop}{\mbox{\tt Propag}}
\newcommand{\Err}{\mbox{\tt Err}}
\newcommand{\EndI}{\mbox{\tt End}}

\newcommand{\Forward}{\mbox{\sf DFS\_F}}
\newcommand{\CoherentD}{\mbox{\sf Coherent\_Done}}
\newcommand{\CoherentV}{\mbox{\sf Coherent\_Verify}}
\newcommand{\CoherentI}{\mbox{\sf Coherent\_Improve}}
\newcommand{\CoherentEI}{\mbox{\sf Coherent\_End}}

\newcommand{\CoherentE}{\mbox{\sf Coherent\_Error}}
\newcommand{\CoherentCycle}{\mbox{\sf CoherentCycle}}
\newcommand{\CoherentTree}{\mbox{\sf CoherentTree}}

\newcommand{\NIC}{\mbox{\sf Succ}}
\newcommand{\pred}{\mbox{\sf Pred}}
\newcommand{\LCandidate}{\mbox{\sf LabCand}}
\newcommand{\Candidate}{\mbox{\sf Candidate}}

\newcommand{\AskI}{\mbox{\sf Ask\_I}}
\newcommand{\AskEI}{\mbox{\sf Ask\_EI}}
\newcommand{\AskV}{\mbox{\sf Ask\_V}}
\newcommand{\AskE}{\mbox{\sf Ask\_E}}
\newcommand{\EndProp}{\mbox{\sf EndPropag}}

\newcommand{\CA}{\mbox{\sf C\_Ancestor}}
\newcommand{\ImproveF}{\mbox{\sf ImproveF}}
\newcommand{\Improve}{\mbox{\sf Improve}}
\newcommand{\TreeEdge}{\mbox{\sf Tree\_Edge}}
\newcommand{\maxLab}{\mbox{\sf MaxLab}}
\newcommand{\MaxC}{\mbox{\sf Max\_C}}
\newcommand{\WayC}{\mbox{\sf Way\_C}}
\newcommand{\Init}{\mbox{\sf Init}}

\newcommand{\Bidule}{\mbox{\sf Nds\_Verify}}

\newcommand{\NdDelete}{\mbox{\sf NdDel}}
\newcommand{\InitV}{\mbox{\sf InitVerify}}

\newcommand{\CDFS}{\mbox{\sf ContinueDFS}}
\newcommand{\EndImprove}{\mbox{\sf End\_Improve}}
\newcommand{\Error}{\mbox{\sf Error}}

\newcommand{\RCA}{\sf R_{D}}
\newcommand{\RCB}{\sf R_{Err}} 
\newcommand{\RDFS}{\sf R_{DFS}}
\newcommand{\RV}{\sf R_{V}}
\newcommand{\RIB}{\sf R_{I}}
\newcommand{\RIE}{\sf R_{E}}

\newcommand{\RDB}{\sf R_{P}}
\newcommand{\RDE}{\sf \bar{R_{P}}}

\newcommand{\LFMST}{\mbox{\sf LoopFreeMST}}

\begin{document}
\title{A New Self-Stabilizing Minimum Spanning Tree Construction with
Loop-free Property}

\author{
L\'{e}lia Blin$^{1,2}$ 
\and
Maria Potop-Butucaru$^{2,3}$
\and
St\'{e}phane Rovedakis$^1$
\and
S\'{e}bastien Tixeuil$^{2,4}$
}

\footnotetext[1]{Universit\'e d'Evry, IBISC, CNRS, France.}
\footnotetext[2]{Univ. Pierre \& Marie Curie - Paris 6, LIP6-CNRS UMR 7606, France.}
\footnotetext[3]{INRIA REGAL, France,}
\footnotetext[4]{INRIA Futurs, Project-team Grand Large.}
\date{\today}

\maketitle
\date{ }

\begin{abstract}
The minimum spanning tree (MST) construction is a classical problem in
Distributed Computing for creating a globally minimized structure
distributedly. Self-stabilization is versatile technique for forward
recovery that permits to handle any kind of transient faults in a
unified manner. The loop-free property provides interesting safety
assurance in dynamic networks where edge-cost changes during operation
of the protocol.

We present a new self-stabilizing MST protocol that improves on
previous known approaches in several ways. First, it makes fewer
system hypotheses as the size of the network (or an upper bound on the
size) need \emph{not} be known to the participants. Second, it is
loop-free in the sense that it guarantees that a spanning tree
structure is always preserved while edge costs change dynamically and
the protocol adjusts to a new MST. Finally, time complexity matches
the best known results, while space complexity results show that this
protocol is the most efficient to date.

\end{abstract}

\thispagestyle{empty}
\setcounter{page}{0}
\newpage 

\section{Introduction}
\label{sec:intro}


Since its introduction in a centralized context~\cite{Prim57,Kruskal56}, the minimum spanning tree (or MST) construction problem gained a benchmark status in distributed computing thanks to the influential seminal work of~\cite{GallagerHS83}. Given an edge-weighted graph $G =(V,E,w)$, where $w$ denotes the edge-weight function, the MST problem consists in computing a tree $T$ spanning $V$, such that $T$ has minimum weight among all spanning trees of $G$.

One of the most versatile technique to ensure forward recovery of distributed systems is that of \emph{self-stabilization}~\cite{D74j,D00b}. A distributed algorithm is self-stabilizing if after faults and attacks hit the system and place it in some arbitrary global state, the system recovers from this catastrophic situation without external (\emph{e.g.} human) intervention in finite time. A recent trend in self-stabilizing research is to complement the self-stabilizing abilities of a distributed algorithm with some additional \emph{safety} properties that are guaranteed when the permanent and intermittent failures that hit the system satisfy some conditions. In addition to being self-stabilizing, a protocol could thus also tolerate a limited number of topology changes~\cite{DH97j}, crash faults~\cite{GP93c,AH93c}, nap faults~\cite{DW97j,PT97j}, Byzantine faults~\cite{DW04j,BDH08c}, and sustained edge cost changes~\cite{CG02j,JT03c}. 

This last property is specially relevant when building spanning trees in dynamic networks, since the cost of a particular edge is likely to evolve through time. If a MST protocol is \emph{only} self-stabilizing, it may adjust to the new costs in such a way that a previously constructed MST evolves into a disconnected or a looping structure (of course, in the abscence of new edge cost changes, the self-stabilization property guarantees that \emph{eventually} a new MST is constructed). Of course, if edge costs change unexpectedly and continuously, a MST can not be maintained at all times. 
Now, a packet routing  algorithm is \emph{loop free}~\cite{Aceves93,Gafni81} if at any point in time the routing tables are free of loops, despite possible modification of the edge-weights in the graph (\emph{i.e.}, for any two nodes $u$ and $v$, the actual routing tables determines a simple path from $u$ to $v$, at any time). The \emph{loop-free} property~\cite{CG02j,JT03c} in self-stabilization guarantees that, a spanning tree being constructed (not necessarily a MST), then the self-stabilizing convergence to a ``minimal'' (for some metric) spanning tree maintains a spanning tree at all times (obviously, this spanning tree is not ``minimal'' at all times). The consequence of this safety property in addition to that of self-stabiization is that the spanning tree structure can still be used (e.g. for routing) while the protocol is adjusting, and makes it suitable for networks that undergo such very frequent dynamic changes.

\paragraph{Related works}
 
Gupta and Srimani~\cite{AntonoiuS97} have presented the first self-stabilizing algorithm for the MST problem. 
It applies on graphs whose nodes have unique identifiers, whose edges have integer edge weights, and a weight can appear at most once in the whole network. To construct the (unique) MST, every node performs the same algorithm. The MST construction is based on the computation of all the shortest paths (for a certain cost function) between all the pairs of nodes. While executing the algorithm, every node stores the cost of all paths from it to all the other nodes. To implement this algorithm, the authors assume that every node knows the number $n$ of nodes in the network, and that the identifiers of the nodes are in $\{1,\dots,n\}$. Every node $u$ stores the weight of the edge $e_{u,v}$ placed in the MST for each node $v\neq u$. Therefore the algorithm requires $\Omega(\sum_{v\neq u}\log w(e_{u,v}))$ bits of memory at node $u$. Since all the weights are distinct integers, the memory requirement at each node is $\Omega(n\log n)$ bits. 
 
Higham and Lyan~\cite{HighamL01} have proposed another self-stabilizing algorithm for the MST problem. 
As~\cite{AntonoiuS97}, their work applies to undirected connected graphs with unique integer edge weights and unique node identifiers, where every node has an upper bound on the number of nodes in the system. The algorithm performs roughly as follows: every edge aims at deciding whether it eventually belongs to the MST or not. For this purpose,  every non tree-edge $e$ floods the network to find a potential cycle, and when $e$ receives its own message back along a cycle, it uses information collected by this message (\emph{i.e.}, the maximum edge weight of the traversed cycle) to decide whether $e$ could potentially be in the MST or not. If the edge $e$ has not received its message back after the time-out interval, it decides to become tree edge. The core memory of each node holds only $O(\log n)$ bits, but the information exchanged between neighboring nodes is of size $O(n \log n)$ bits, thus only slightly improving that of \cite{AntonoiuS97}.

To our knowledge, \emph{none} of the self-stabilizing MST construction protocols is loop-free. Since the aforementioned two protocols also make use of the knowledge of the global number of nodes in the system, and assume that no two edge costs can be equal, these extra hypoteses make them suitable for static networks only.

Relatively few works investigate merging self-stabilization and loop free routing, with the notable exception of~\cite{CG02j,JT03c}. While~\cite{CG02j} still requires that a upper bound on the network diameter is known to every participant, no such assumption is made in~\cite{JT03c}. Also, both protocols use only a reasonable amount of memory ($O(\log n)$ bits per node). However, the metrics that are considered in~\cite{CG02j,JT03c} are derivative of the shortest path (\emph{a.k.a.} SP) metric, that is considered a much easier task in the distributed setting than that of the MST, since the associated metric is \emph{locally optimizable}~\cite{GS99c}, allowing essentially locally greedy approaches to perform well. By contrast, some sort of \emph{global optimization} is needed for MST, which often drives higher complexity costs and thus less flexibility in dynamic networks.

\paragraph{Our contributions} 

We describe a new self-stabilizing algorithm for the MST problem. Contrary to previous self-stabilizing MST protocols, our algorithm does not make any assumption about the network size (including upper bounds) or the unicity of the edge weights. Moreover, our solution improves on the memory space usage since each participant needs only $O(\log n)$ bits, and node identifiers are not needed.

In addition to improving over system hypotheses and complexity, our algorithm provides additional safety properties to self-stabilization, as it is loop-free. Compared to previous protocols that are both self-stabilizing and loop-free, our protocol is the first to consider non-monotonous tree metrics.

\begin{table}[t]
\begin{center}
\scalebox{1}
{
\begin{tabular}{|l|c|c|c|c|c|}
\hline
 & metric & size known & unique weights & memory usage &  loop-free \\
\hline
 \cite{AntonoiuS97}& \textbf{MST} & yes & yes &$\Theta(n \log n)$ & no\\
 \cite{HighamL01}& \textbf{MST} & upper bound & yes &$\Theta(n \log n)$& no\\
 \cite{CG02j}& SP & upper bound & \textbf{no} & $\mathbf{\Theta(\log n)}$& \textbf{yes} \\
 \cite{JT03c}& SP & \textbf{no} & \textbf{no} & $\mathbf{\Theta(\log n)}$& \textbf{yes} \\
\hline
This paper & \textbf{MST} & \textbf{no} & \textbf{no} & $\mathbf{\Theta(\log n)}$ & \textbf{yes}\\
\hline
\end{tabular}
}
\caption{\small Distributed Self-Stabilizing algorithms for the MST and loop-free SP problems}
\label{tableresume}
\end{center}
\end{table}

The key techniques that are used in our scheme include fast construction of a spanning tree, that is continuously improved by means of a pre-order construction over the nodes. The cycles that are considered over time are precisely those obtained by adding one edge to the evolving spanning tree. Considering solely that type of cycles reduces the memory requirement at each node compared to \cite{AntonoiuS97,HighamL01} because the latter consider all possible paths  connecting pairs of nodes. Moreover, constructing and using a pre-order on the nodes allows our algorithm to proceed in a completely asynchronous manner, and without any information about the size of the network, as opposed to~\cite{AntonoiuS97,HighamL01}. The main characteristics of our solution are presented in Table~\ref{tableresume}, where a boldface denotes the most useful (or efficient) feature for a particular criterium.

\section{Model and notations} 
\label{sec:model}


We consider an undirected weighted connected network $G=(V,E,w)$ where $V$ is the set of nodes, $E$ is the set of edges and $w: E \rightarrow {\mathbb R^+}$ is a positive cost function. 
Nodes represent processors and edges represent bidirectional communication links. Additionally, we consider that $G=(V,E,w)$ is a network in which the weight of the communication links may change value. 
We consider anonymous networks (i.e., the processor have no IDs), with one distinguished node, called the \emph{root}\footnote{Observe that the two self-stabilizing MST algorithms mentioned in the Previous Work section assume that the nodes have distinct IDs with no distinguished  nodes. Nevertheless, if the nodes have distinct IDs then it is possible to elect one node as a leader in a self-stabilizing manner. Conversely, if there exists one distinguished node in an anonymous network, then it is possible to assign distinct IDs to the nodes in a self-stabilizing manner~\cite{Dolev00}. Note that it is not possible to compute deterministically  a MST in a fully anonymous network (i.e., without any distinguished node), as proved in \cite{AntonoiuS97}.}. Throughout the paper, the root is denoted $r$. We denote by $\deg(v)$ the number of $v$'s neighbors in $G$. The $\deg(v)$ edges incident to any node $v$ are labeled from 1 to $\deg(v)$, so that a processor can distinguish the different edges incident to a node. 

The processors asynchronously execute their programs consisting of a set of variables and a finite set of rules. The variables are part of the shared register which is used to communicate with the neighbors. A processor can read and write its own registers and can read the shared registers of its neighbors. 
Each processor executes a program consisting of a sequence of guarded rules. Each \emph{rule} contains a \emph{guard} (boolean expression over the variables of a node and its neighborhood) and an \emph{action} (update of the node variables only). Any rule whose guard is \emph{true} is said to be \emph{enabled}. A node with one or more enabled rules is said to be \emph{privileged} and may make a \emph{move} executing the action corresponding to the chosen enabled rule.

A {\it local state} of a node is the value of the local variables of the node and the state of its program counter. A {\it configuration} of the system $G=(V,E)$ is the cross product of the local states of all nodes in the system. The transition from a configuration to the next one is produced by the execution of an action at a node. A {\it computation} of the system is defined as a \emph{weakly fair, maximal} sequence of configurations, $e=(c_0, c_1, \ldots c_i, \ldots)$, where each configuration $c_{i+1}$ follows from $c_i$ by the execution of a single action of at least one node. During an execution step, one or more processors execute an action and a processor may take at most one action. \emph{Weak fairness} of the sequence means that if any action in $G$ is continuously enabled along the sequence, it is eventually chosen for execution. \emph{Maximality} means that the sequence is either infinite, or it is finite and no action of $G$ is enabled in the final global state.

In the sequel we consider the system can start in any configuration. That is, the local state of a node can be corrupted. Note that we don't make any assumption on the bound of corrupted nodes. In the worst case all the nodes in the system may start in a corrupted configuration. In order to tackle these faults we use self-stabilization techniques.

\begin{definition}[self-stabilization]
Let $\mathcal{L_{A}}$ be a non-empty \emph{legitimacy predicate}\footnote{A legitimacy predicate is defined over the configurations of a system and is an indicator of its correct behavior.} of an algorithm $\mathcal{A}$ with respect to a specification predicate $Spec$ such that every configuration satisfying $\mathcal{L_{A}}$ satisfies $Spec$. Algorithm $\mathcal{A}$ is \emph{self-stabilizing} with respect to $Spec$ iff the following two conditions hold:\\
\textsf{(i)} Every computation of $\mathcal{A}$ starting from a configuration satisfying $\mathcal{L_A}$ preserves $\mathcal{L_A}$ (\emph{closure}).  \\
\textsf{(ii)} Every computation of $\mathcal{A}$ starting from an arbitrary configuration contains a configuration that satisfies $\mathcal{L_A}$ (\emph{convergence}).
\end{definition}

We define bellow a \emph{loop-free} configuration of a system as a configuration which contains paths with no cycle between any couple of nodes in the system.


\begin{definition}[Loop-Free Configuration]
Let $Cycle(u,v)$ be the following predicate defined for two nodes $u,v$ on configuration $C$, with $P(u,v)$ a path from $u$ to $v$ described by $C$:
$$Cycle(u,v) \equiv \exists P(u,v), P(v,u) : P(u,v) \cap P(v,u) = \emptyset.$$
A loop-free configuration is a configuration of the system which satisifes $\forall u,v: Cycle(u,v)=false$.
\end{definition}

%

We use the definition of a loop-free configuration to define a \emph{loop-free stabilizing} system.


\begin{definition}[Loop-Free Stabilization]
A distributed system is called loop-free stabilizing if and only if it is self-stabilizing and there exists a non-empty set of configurations such that the following conditions hold: \textsf{(i)} Every execution starting from a loop-free configuration reaches a loop-free configuration (closure). \textsf{(ii)} Every execution starting from an arbitrary configuration contains a loop-free configuration (convergence).
\end{definition}

In the sequel we study the loop-free self-stabilizing \LFMST problem.
The legitimacy predicate $\mathcal{L_{A}}$ for the \LFMST problem is the conjunction of the following two predicates:
\textsf{(i)} a tree $T$ spanning the network is constructed. \textsf{(ii)} $T$ is a minimum spanning tree of $G$ (i.e., 
$\forall T', W(T) \leq W(T')$, with $T'$ be a spanning tree of $G$ and $W(S)=\sum_{e \in S} w(e)$ be the cost of the subgraph $S$).

\section{The Algorithm \LFMST}

In this section, we describe our self-stabilizing algorithm for the MST problem. We call this algorithm \LFMST. In the next section, we shall prove the correctness of this algorithm, and demonstrate that it satisfies all the desired properties listed in Section~\ref{sec:intro}, including the loop-freedomness property. Let us begin by an informal description of  \LFMST\/ aiming at underlining its main features. 

\subsection{High level description}
\label{sec:high}

\LFMST\/ is based on the red rule. That is, for constructing a MST, the algorithm successively deletes the edges of maximum weight within every cycle. For this purpose,  a spanning tree is maintained, together with a pre-order labeling of its nodes. Given the current spanning tree $T$ maintained by our algorithm, every edge $e$ of the graph that is not in the spanning tree creates an unique cycle in the graph when added to $T$. This cycle is called \emph{fundamental cycle}, and is denoted by $C_e$. (Formally, this cycle depends on $T$; Nevertheless no confusion should arise from omitting $T$ in the notation of $C_e$). If $w(e)$ is not the maximum weight of all the edges in $C_e$, then, according to the red rule, our algorithm swaps $e$ with the edge $f$ of  $C_e$ with maximum weight . This swapping procedure is called an \emph{improvement}.  A straightforward consequence of the red rule is that if no improvements are possible then the current spanning tree is a minimum one.

Algorithm \LFMST\/ can be decomposed in three procedures:  
\vspace*{-0,3cm}
\begin{itemize}
\item Tree construction \vspace*{-0,3cm}
\item Token label circulation \vspace*{-0,3cm}
\item Cycle improvement \vspace*{-0,3cm}
\end{itemize}
The latter procedure (Cycle improvement) is in fact the core of our contribution. Indeed, the two first procedures are simple modifications of existing self-stabilizing algorithms, one for building a spanning tree, and the other for labelling its nodes. We will show how to compose the original procedure "Cycle improvement" with these two existing procedures. Note that "Cycle improvement" differs from the previous self-stabilizing implementation of the improvement swapping in~\cite{HighamL01} by the fact that it does not require any a priori knowledge of the network, and it is loop-free. 

\LFMST\/ starts by constructing a spanning tree of the graph, using the self-stabilizing loop-free algorithm "Tree construction" described in~\cite{JT03}. The two other procedures are performed concurrently. A token circulates along the edges of the current spanning tree, in a self-stabilizing manner. This token circulation uses algorithms proposed in \cite{DGPV01,PetitV07} as follows. A non-tree-edge can belong to at most one fundamental cycle, but a tree-edge can belong to several fundamental cycles. Therefore, to avoid simultaneous possibly conflicting improvements, our algorithm considers the cycles in order. For this purpose, the token labels the nodes of the current tree in a DFS order (pre-order). This labeling is then used to find the unique path between two nodes in the spanning tree in a distributed manner, and enables computing the fundamental cycle resulting from adding one edge to the current spanning tree. 

\begin{figure}[!h]
\begin{center}
\includegraphics[scale=0.5]{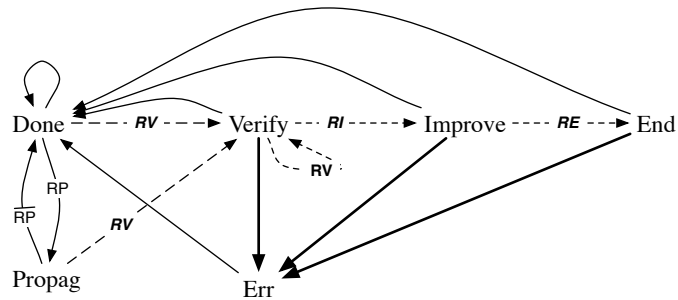}
\end{center}
\caption{Evolution of the node's state in cycle improvement module. Rule $\RCA$ is depicted in plain. Rule $\RCB$ is depicted in bold.}
\label{fig:etats}
\end{figure}

\begin{figure}[!h]
\begin{center}
\includegraphics[scale=0.3]{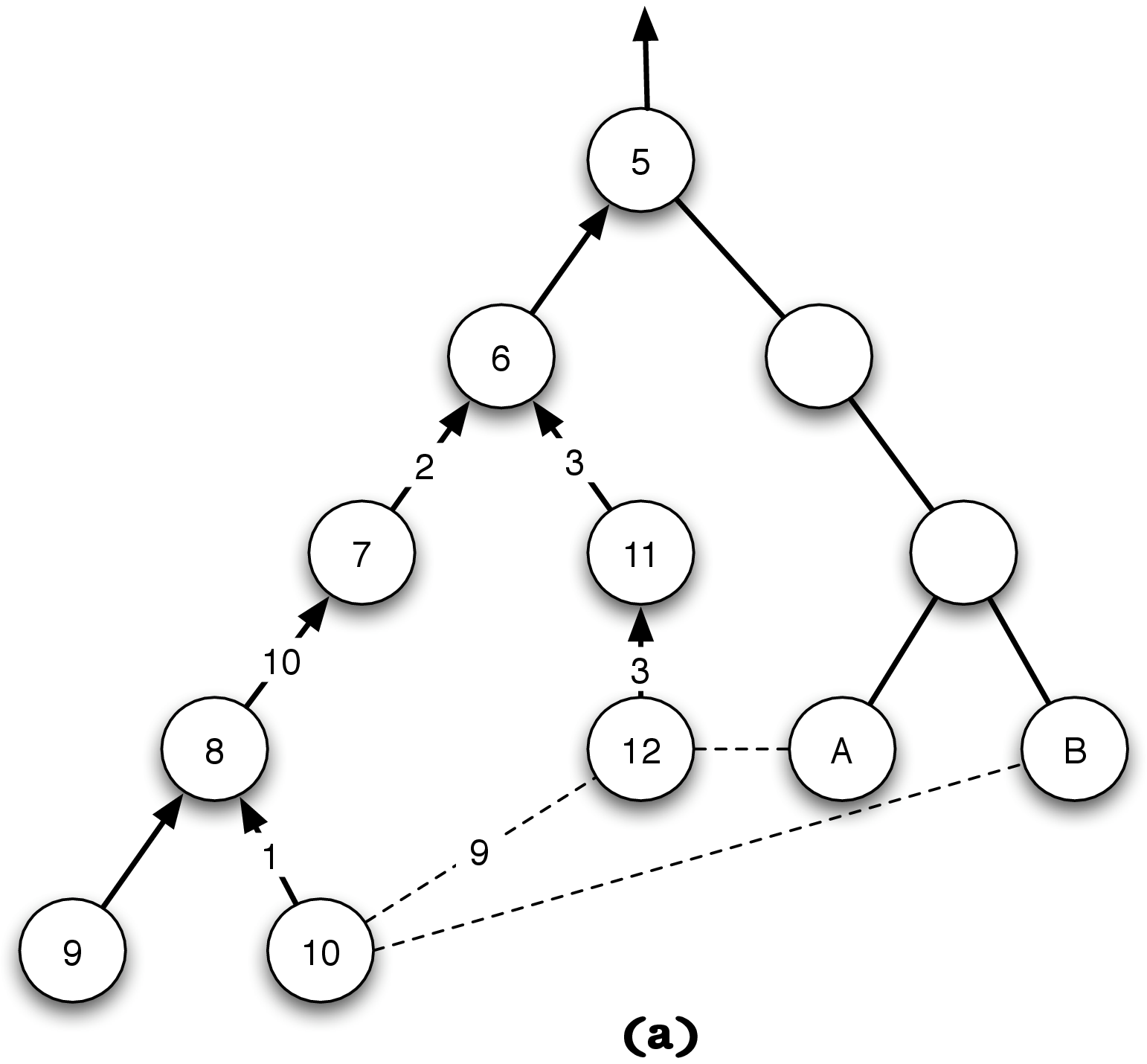}
\includegraphics[scale=0.3]{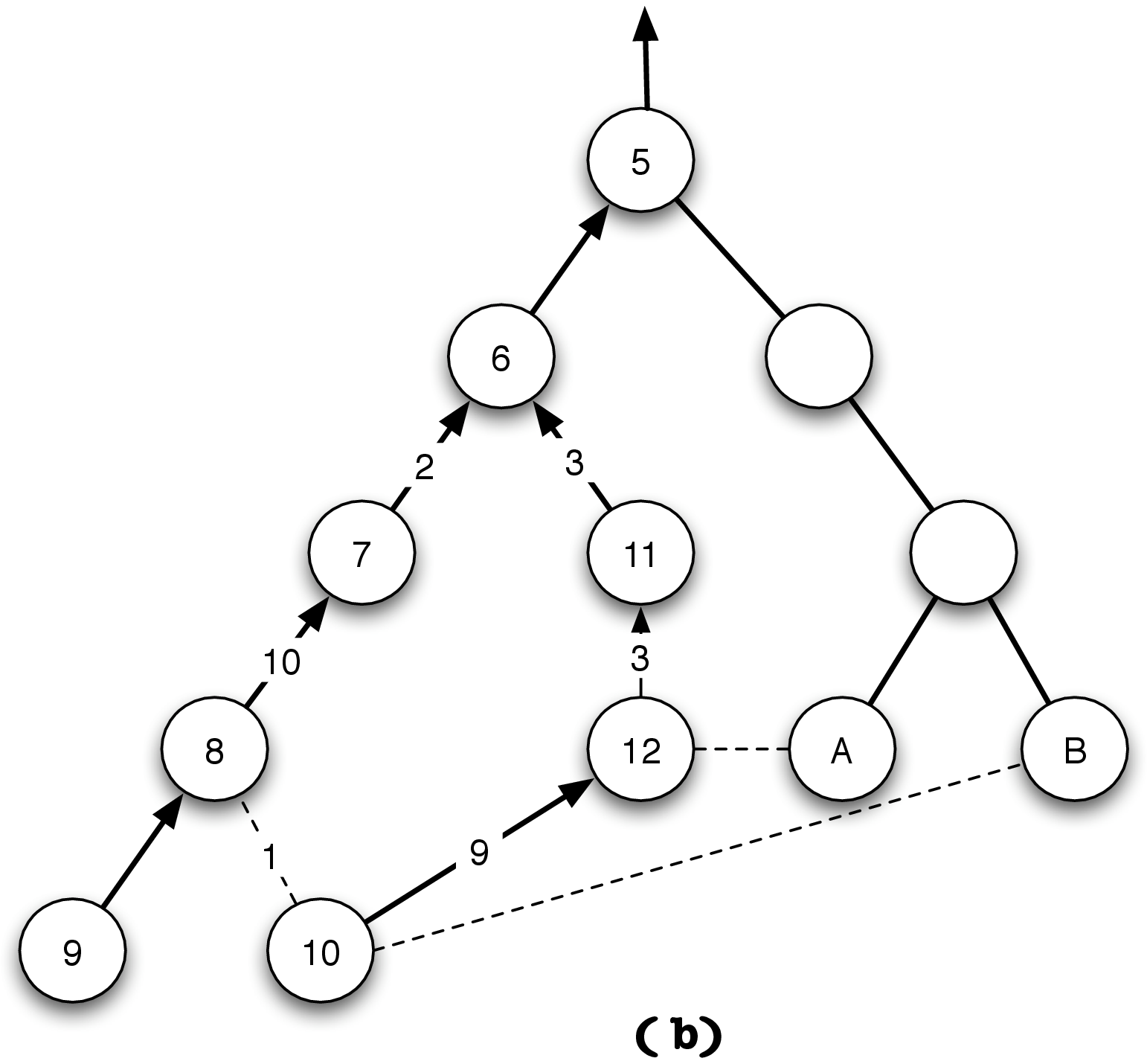}
\includegraphics[scale=0.3]{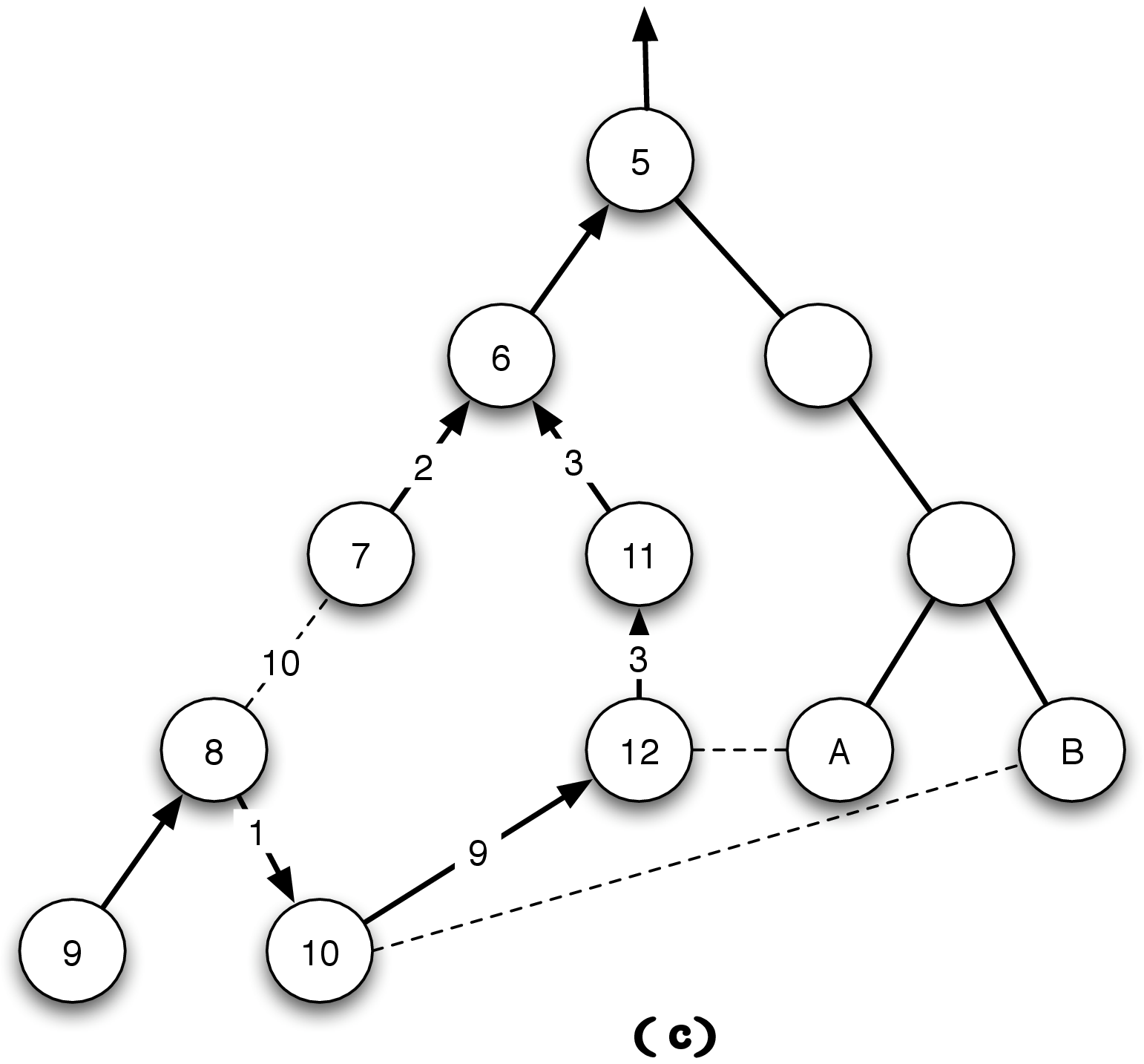}
\end{center}
\caption{Example of a loop-free improvement of the current spanning tree. The direction of the edges indicate the parent relation. Edges in the spanning tree are depicted as plain lines; Edges not in the spanning tree are denoted by dotted lines. }
\label{fig:ex1}
\end{figure}

We now sketch the description of the procedure "Cycle improvement" (see Figure~\ref{fig:etats}). When the token arrives at a node $u$ in a state \Do, it checks whether $u$ has some incident edges not in the current spanning tree $T$ connecting $u$ with some other node $v$ with smaller label.  If it is the case, then enters state \Ve. Let $e=\{u,v\}$. Node $u$ then initiates a traversal of the fundamental cycle $C_e$ for finding the edge $f$ with maximum weight in this cycle. If $w(f)=w(e)$ then no improvement is performed. Else an improvement is possible, and $u$ enters State \Imp. Exchanging $e$ and $f$ in $T$ results in a new tree $T'$. The key issue here is to perform this exchange in a loop-free manner. Indeed, one cannot be sure that two modifications of the current tree (i.e., removing $f$ from $T$, and adding $e$ to $T$) that are applied at two distant nodes will occur simultaneously. And if they do not occur simultaneously, then there will a time interval during which the nodes will not be connected by a spanning tree. Our solution for preserving loop-freedomless relies on a sequence of successive local and atomic changes, involving a single variable. This variable is a pointer to the current parent of a node in the current spanning tree. To get  the flavor of our method, let us consider the example depicted on Figure~\ref{fig:ex1}.  In this example, our algorithm has to exchange the edge $e=\{10,12\}$ of weight 9, with the edge $f=\{7,8\}$ of weight 10 (Figure~\ref{fig:ex1}(a)). Currently, the token is at node $12$. The improvement is performed in two steps, by a sequence of two local changes.  First, node 10 switches  its parent from 8 to 12 (Figure~\ref{fig:ex1}(b)). Next, node 8 switches its parent from  7 to 10 (Figure~\ref{fig:ex1}(c)). A spanning tree is preserved at any time during the execution of these changes. 

Note that any modification of the spanning tree makes the current labeling globally inaccurate, i.e., it is not necessarily a pre-order anymore. However, the labeling remains a pre-order in the portion of the tree involved in the exchange. For instance, consider again the example depicted on Figure~\ref{fig:ex1}(c). When the token will eventually reach node $A$, it will label it by some label $\ell>12$. The exchange of $e=\{10,12\}$ and $f=\{7,8\}$ has not changed the pre-order for the fundamental cycle including edge $\{A,12\}$.  However, when the token will eventually reach node $B$ and label it $\ell'>\ell$, the exchange of $e=\{10,12\}$ and $f=\{7,8\}$ has changed the pre-order for the fundamental cycle including edge $\{B,9\}$: the parent of node labeled $10$ is labeled $12$ whereas it should have a label smaller than $10$ in a pre-order.  When the pre-order is modified by an exchange, the inaccurately  labeled node changes its state to  \Err, and stops the traversal of the fundamental cycle. The token is then informed that it can discard this cycle, and carry on the traversal of the tree. 

\subsection{Detailed level description}

We now enter into the details of Algorithm  \LFMST. First, let us state all variables used by the algorithm. Later on, we will describe its predicates and its rules.

\paragraph{Variables}

For any node $v \in V(G)$, we denote by $N(v)$ the set of all neighbors of $v$ in $G$. Algorithm \LFMST\/ maintains the set $N(v)$ at every node $v$. We use the following notations: 

\begin{itemize}
\item $\parent_v$: the parent of $v$ in the current spanning tree;\vspace*{-0,3cm}
\item $\La_v$:  the integer label assigned to $v$;\vspace*{-0,3cm}
\item $\dist_v$: the distance (in hops) from $v$ to the root in the current spanning tree;\vspace*{-0,3cm}
\item $\state_v$: the state of node $v$, with values in $\{\mbox{\Do, \Ve, \Imp, \EndI, \Prop, \Err}\}$;\vspace*{-0,3cm}
\item $\DefC_v$: the pair of labels of the two extremities of the non tree-edge corresponding to the current fundamental cycle.\vspace*{-0,3cm}
\item $\VarC_v$: a pair of variables: the first one is the maximum edge-weight in  the current fundamental cycle;  the second one is a  (boolean) variable in $\{{\mbox{\Be, \Af}\}}$;\vspace*{-0,3cm}
\item $\next_v$: the successor of $v$ in the current fundamental cycle.
\end{itemize}

\paragraph{Consistency rules}

The first task executed by \LFMST\/ is to check the consistency of the variables of each node; See Figure~\ref{fig:etats}.  
\Do\/ is the standard state of a node when this node has not the token, or is not currently visited by the traversal of a fundamental cycle. When the variables of a node are detected to be not coherent, the state of the node becomes \Err\/  thanks to rule $\RCB$.  There is one predicate in $\RCB$ for each state, except for state $\Prop$, to check whether the variables of the node are consistent (see Figure \ref{fig:correct_predicates}). The rule $\RCA$ allows the node to return to the standard state \Do. More precisely, rule $\RCA$ resets the variables, and stops the participation of the node to any improvement.

\begin{description}
\item[$\RCB$: (Bad label)]~\\
\textbf{If} $\CoherentCycle(v) \wedge \Error(v) \wedge \DefC[0]_v \neq \La_v \wedge \EndProp(v)$ \textbf{then} $\state_v:=\Err;$ 
\item[$\RCA$: (Improvement consistency)]~\\ \textbf{If} $\neg \CoherentCycle(v) \wedge \EndProp(v)$\\
\textbf{then} $\state_v:=\Do; \DefC_v:=(\La_v,\done); \VarC_v:=(0,\Be); \next_v:=\emptyset$; \\
\end{description}

\begin{figure}[!ht]
\fbox{
\begin{minipage}{17cm}
\footnotesize{
\begin{description}
\item[$\CoherentCycle(v)$] $\equiv \CoherentD(v) \vee \CoherentV(v) \vee \CoherentI(v) \vee \CoherentEI(v) \vee \CoherentE(v)$
\item[$\CoherentD(v)$]$\equiv \state_v=\Do \wedge \next_v=\emptyset \wedge \DefC_v=(\La_v,\done)   \wedge \VarC_v=(0,\Be)$
\item[$\CoherentV(v)$]$\equiv \state_v=\Ve \wedge \next_v=\NIC(v)  \wedge [(\Init(v) \wedge \VarC_x=(0,\Be)) \vee \Bidule(v) ]$
\item[$\CoherentI(v)$\hspace*{-0,3cm}] $\equiv \state_v=\Imp \wedge \next_v=\NIC(v) \wedge  \DefC_v=\DefC_{\parent_v} \wedge \VarC_v=\VarC_{\parent_v}$
\item[$\CoherentEI(v)$]$\equiv \state_v=\EndI \wedge \DefC_v=\DefC_{\parent_v} \wedge  (\NdDelete(v)  \vee \AskEI(v))$
\item[$\CoherentE(v)$]$\equiv \state_v=\Err \wedge (\next_v=\NIC(v)=\emptyset \vee \AskE(v)) \wedge \DefC_v=\DefC_{\pred(v)}$
\item[$\CoherentTree(v)$] \hspace*{-0,3cm} $\equiv$  \hspace*{-0,1cm}$(v=r \wedge \dist_v=0 \wedge st_v=N) \vee (v \neq r \wedge Safe_v \wedge rw_v=\dist_v) \vee \state_{\parent_v}=\Imp \vee \state_{\parent_v}=\Prop$
\item[$\AskV(v)$]$ \equiv  \state_{\pred(v)}=\Ve$
\item[$\AskI(v)$\hspace*{-0,2cm}]$\equiv (\state_{\pred(v)}=\Imp \wedge \VarC[1]_{\pred(v)}=\Be) \vee (\state_{\next_v}=\Imp \wedge \VarC[1]_{\next_v}=\Af)$
\item[$\AskEI(v)$]$ \equiv  (\exists u \in N(v), \parent_u=v \wedge \state_u=\EndI \wedge \DefC_u=\DefC_v)$
\item[$\AskE(v)$]$ \equiv \next_v \neq \emptyset \wedge \state_{\next_v}=\Err \wedge \DefC_v=\DefC_{\next_v}$
\end{description}
}
\end{minipage}
}
\caption{Corrections predicates used by \LFMST.}
\label{fig:correct_predicates}
\end{figure}

\begin{figure}[!ht]
\fbox{
\begin{minipage}{16cm}
\footnotesize{
\begin{description}
\item[$\TreeEdge(v,u)$]$ \equiv \parent_v=u \vee \parent_u=v$
\item[$\CA(v)$] $\equiv \parent_v \neq \next_v \wedge \parent_v \neq \pred(v)$
\item[$\Init(v)$] $\equiv \Forward(v) \wedge \DefC[0]_v=\La_v$
\item[$\Bidule(v)$]$\equiv [(\AskV(v)  \wedge \VarC_v=(\MaxC(v),\WayC(v))) \vee \AskI(v)] \wedge  \DefC_v=\DefC_{\pred(v)}$
\item[$\NdDelete(v)$]$\equiv  \state_{\parent_v} \neq \Do \wedge \state_{\parent_v} \neq \Prop \wedge \neg \Improve(v)$
\end{description}
}
\end{minipage}
}
\caption{Corrections predicates used by the algorithm.}
\label{fig:correct_predicates3}
\end{figure}

\paragraph{Tree construction}
\label{subsubsec:treeconst}

\LFMST\/ starts by constructing a spanning tree of the graph, using the self-stabilizing loop-free algorithm "Tree construction" described in~\cite{JT03}. This algorithm constructs a BFS, and uses two variables $parent$ and $distance$. During the execution of our algorithm, these two variables are subject to the same rules as in~\cite{JT03}. After each modification of the spanning tree, the new distance to the parent is propagated in sub-trees by Rules  $\RDB$ and $\RDE$.

\begin{description}
\item[$\RDB$: (Distance propagation)]~\\
\footnotesize{
\textbf{If} $\CoherentD(v) \wedge \neg \AskV(v) \wedge (\state_{\parent_v}=\Imp \vee \state_{\parent_v}=\Prop) \wedge \next_v \neq \parent_v \wedge \pred(v) \neq \parent_v \wedge \dist_v \neq \dist_{\parent_v}+1$\\
\textbf{then} $\state_v:=\Prop; \dist_v:=\dist_{\parent_v}+1;$}
\item[$\RDE$: (End distance propagation)]~\\
\footnotesize{
\textbf{If} $\state_v=\Prop \wedge \EndProp(v)$\\
\textbf{then} $\state_v:=\Do; \DefC_v:=(\La_v,\done); \VarC_v:=(0,\Be); \next_v:=\emptyset;$}
\end{description}

\paragraph{Token circulation and pre-order labeling}

\LFMST\/ uses the algorithm described in \cite{DGPV01} to provide each node $v$ with a label $\La_v$. Each label is unique in the network traversed by the token. This labeling is used to find the unique path between two nodes in the spanning tree, in a distributed manner. For this purpose, we use the snap-stabilizing algorithm described in \cite{PetitV07} for the circulation of a token in the spanning tree. We have slightly modified this algorithm because \LFMST\/  stops the token circulation at a node during  the "Cycle improvement" procedure. A node $v$ knows if it has the token by applying predicate $\Init(v)$.  Rule $\RDFS$ guides the circulation of the token. The token carries on its tree traversal if one of  the following three conditions is satisfied: (i) there is no improvement which could be initiated by the node which holds the token, (ii) an improvement was performed in the current cycle, or (iii) inconsistent node labels were detected in the current cycle. The latter is under the control of  Predicate $\CDFS(v)$. 

\begin{description}
\item[$\RDFS$: (Continue DFS token circulation)]~\\ \textbf{If} $\CoherentCycle(v) \wedge \Init(v) \wedge \CDFS(v)$\\ 
\textbf{then} $\state_v:=\Do; \DefC[1]_v=\done;$
\end{description}

\paragraph{Cycle improvement rules}

The procedure "Cycle improvement" is the core of \LFMST. Its role is to avoid disconnection of the current spanning tree, while successively improving the tree until reaching a MST.  The procedure can be decomposed in four tasks:  (1)  to check whether the fundamental cycle of the non-tree edge  has an improvement or not, (2) perform the improvement if any, (3) update the distances, and (4) resume the token circulation.  

Let  us start by describing the first task. A node $u$ in state $\Do$ changes its state to $\Ve$ if its variables are in consistent state, it has a token, and it has identified a candidate (i.e., an incident non-tree edge $e= \{u,v\}$ whose other extremity $v$ has a smaller label than the one of $u$). 
The latter is under the control of Predicate $\InitV(v)$, and the  variable $\VarC_v$ contains the label of $u$ and $v$. If the three conditions are satisfied, then the verification of the fundamental cycle $C_e$ is initiated from node $u$, by applying rule $\RV$. The goal of this verification is twofold: first, to verify whether $C_e$ exists or not, and, second, to save information about the maximum edge weight and the location of the edge of maximum weight in $C_e$. These information are stored in the variable $\WayC(v)$. In order to respect the orientation in the current spanning tree,  the node $u$ or $v$ that initiates the improvement depends on the localization of the maximum weight edge $f$ in $C_e$. More precisely, let $r$ be the least common ancestor of nodes $u$ and $v$ in the current tree. If $f$ occurs before $r$ in $T$ in the traversal of $C_e$ from $u$ starting by edge $(u,v)$, then the improvement starts from $u$, otherwise the improvement starts from $v$. To get the flavor of our method, let us consider the example depicted on Figure~\ref{fig:ex1}. In this example, $f$ occurs after the least common ancestor (node 6). Therefore node 10 atomically swaps its parent to respect the orientation. However, if one replaces in the same example the weight of edge  $\{11,6\}$ by 11 instead of 3, then $f$ would occur before $r$, and thus node 12 would have to atomically swaps its parent. The relative places of $f$ and $r$ in the cycle is indicated by Predicate $\WayC(v)$ that returns two different values: $\Be$ or $\Af$.  
During the improvement of the tree,  the fundamental cycle is modified. It is crucial to save information about this cycle during this modification. In particular, the successor of a node $w$ in a cycle, stored in the variable $\next_w$, must be preserved. Its value is computed by Predicate $\NIC(v)$ which uses node labels to identify the current examined fundamental cycle. Each node is able to compute its predecessor in the fundamental cycle by applying Predicate $\pred(v)$. The state of a node is compared with the ones of its successor and predecessor to detect potential inconsistent values. At the end of this task, the node $u$ learns the maximum weight of the cycle $C_e$ and can decide whether it is possible to make an improvement or not. If not, but there is another non-tree edge $e'$ that is candidate for potential replacement, then $u$ verifies $C_{e'}$. Otherwise the token carries on its traversal, and rule $\RDE$ is applied. 

\begin{description}
\item[$\RV$:] (Verify rule)~\\ 
\footnotesize{
\textbf{If} $\CoherentCycle(v) \wedge \neg \Error(v) \wedge (\InitV(v) \vee [\neg \Init(v) \wedge (\CoherentD(v) \vee \state_v=\Prop) \wedge \AskV(v)])$\\
\textbf{then} $\state_v:=\Ve;$\\
\hspace*{0,7cm} \textbf{If} $\Forward(v)$ \textbf{then} $\DefC[1]_v:=\LCandidate(v);$\\
\hspace*{0,7cm} \textbf{Else} $\DefC_v:=\DefC_{\pred(v)}; \VarC_v:=(\MaxC(v),\WayC(v));\next_v:=\NIC(v);$
}
\end{description}

\begin{figure}[!htp]
\fbox{
\begin{minipage}{17cm}
\footnotesize{
\begin{description}
\item[$\pred(v)$]$ \equiv \arg \min\{\La_u: u \in N(v) \wedge \state_u \neq \Do \wedge \state_u \neq \Prop \wedge \next_u=v \}$  if $u$ exists, $\emptyset$ otherwise
\item[$\maxLab(v,x)$]$\equiv \arg \max \{\La_s: s \in N(v) \wedge \La_s<x \}$
\item[$\NIC(v)$]$ \equiv \left\{ \begin{array}{ll} \VarC[0]_v & \mbox{if } \DefC[1]_v=\La_v \\ \parent_v & \mbox{if } (\La_v>\DefC[1]_v \wedge \state_v=\Ve) \vee (\La_v<\DefC[1]_v \wedge \\ & (\state_v=\Imp \vee \state_v=\EndI))\\  \maxLab(v,\DefC[1]_v) & \mbox{if } (\La_v<\DefC[1]_v \wedge \state_v=\Ve)\\   \maxLab(v,\La_v) &\mbox{if }(\La_v>\DefC[1]_v \wedge (\state_v=\Imp \vee \state_v=\EndI)) \end{array} \right.$
\item[$\MaxC(v)$]$ \equiv \max\{\VarC[0]_{\pred(v)}, w(v,\pred(v))\}$
\item[$\WayC(v)$]$ \equiv \left\{ \begin{array}{ll} \mbox{\Af} & \mbox{if }  \VarC[0]_v\neq\VarC[0]_{\pred(v)} \wedge \La_v >\La_{\pred(v)}\\ \VarC[1]_{\pred(v)} & \mbox{otherwise} \end{array} \right.$
\item[$\LCandidate(v)$]\hspace*{-0,2cm}$ \equiv \min\{\La_u: u \in N(v) \wedge \La_u<\La_v \wedge \neg \TreeEdge(v,u) \wedge \La_u \succ \DefC[1]_v\}\footnote{$\succ$ order on neighbor labels for which '\End' is the biggest element and '\done' is the smallest one.}$ if $u$ exists, $\End$ otherwise
\end{description}
}
\end{minipage}
}
\caption{Predicates used by the algorithm.}
\label{fig:algo_predicates1}
\end{figure}

If $C_e$ can yield an improvement, then rule $\RIB$ is executed. By this rule, a node enters in state \Imp, and changes its parent to its predecessor  if $\VarC[1]_v=\Be$ (respectively to its successor if $\VarC[1]_v=\Af$). For this purpose, it uses the variable $\next_v$ and the predicate $\pred(v)$ . 

\begin{description}
\item[$\RIB$: (Improve rule)]~\\
\footnotesize{
\hspace*{-1cm}\textbf{If} $\CoherentCycle(v) \wedge \neg \Error(v) \wedge \CoherentV(v) \wedge \Improve(v) \wedge\neg \CA(v) \wedge [(\Forward(v) \wedge \AskV(v))  \vee \AskI(v)]$\\
\hspace*{-1cm}\textbf{then}  $\state_v:=\Imp;$\\
\hspace*{0,cm} \textbf{If} $\Forward(v) \vee \state_{\pred(v)}=\Imp$ \textbf{then} $\VarC_v:=\VarC_{\pred(v)}  $\\ 
\hspace*{0,cm} \textbf{If} $(\Forward(v) \wedge \VarC[1]_v=\Be) \vee \neg \Forward(v)$ \textbf{then} $\parent_v:=\pred(v);$\\
\hspace*{0,cm} \textbf{If} $\state_{\next_v}=\Imp$ \textbf{then} $\VarC_v:=\VarC_{\next_v} ;  \parent_v:=\next_v; $\\
\hspace*{0,cm} \textbf{If} $w(v,\next_v) \geq \VarC[0]_v \textbf{ then } \next_v=\NIC(v)$\\
\hspace*{0,cm} $\dist_v:=\dist_{\parent_v}+1;$}
\end{description}

At the end of an improvement, it is necessary to inform the node holding the token that it has to carry on its traversal. This is the role of rule $\RIE$. It is also necessary to inform all nodes impacted by the modification that they have to update their distances to the root (see Section~\ref{subsubsec:treeconst}). 

\begin{description}
\item[$\RIE$: (End of improvement rule)]~\\
\footnotesize{
\textbf{If} $\CoherentCycle(v) \wedge \neg \Error(v) \wedge \EndImprove(v) \wedge \EndProp(v)$\\
\textbf{then} $\state_v:=\EndI;$}
\end{description}

%

\begin{figure}[!htp]
\fbox{
\begin{minipage}{16cm}
\footnotesize{
\begin{description}
\item[$\Candidate(v)$]$ \equiv \LCandidate(v) \neq \End $
\item[$\InitV(v)$]$ \equiv \Init(v) \wedge \Candidate(v) \wedge (\CoherentD(v) \vee [\CoherentV(v) \wedge \neg \Improve(v) \wedge \neg \CA(v) \wedge \AskV(v)])$
\item[$\ImproveF(v,x)$] $\equiv \neg \TreeEdge(v,x)) \wedge \max(\VarC[0]_v,\VarC[0]_x)  >  w(v,x)$
\item[$\Improve(v)$] $\equiv \ImproveF(v,\pred(v)) \vee  \ImproveF(v,\next_v) $
\item[$\EndImprove(v)$] $\equiv  \CoherentI(v) \wedge (\NdDelete(v) \vee \AskEI(v))$
\item[$\CDFS(v)$] $\equiv (\Init(v) \wedge [([\CoherentD(v) \vee (\CoherentV(v) \wedge \neg \ImproveF(v,\pred(v)) \wedge \AskV(v))] \wedge \neg \Candidate(v)) \vee \CoherentEI(v) \vee \Error(v)]) \vee \neg \Forward(v)$
\item[$\Error(v)$]$\equiv \state_v \neq \Do \wedge \state_v \neq \Err \wedge (\next_v=\NIC(v)=\emptyset \vee \AskE(v))$
\item[$\EndProp(v)$] $\equiv (\forall u \in N(v), \parent_u=v \wedge \state_u=\Do \wedge \dist_u=\dist_v+1)$
\end{description}
}
\end{minipage}
}
\caption{Predicates used by the algorithm.}
\label{fig:algo_predicates}
\end{figure}

\paragraph{Module composition}

\begin{figure}[!h]
\begin{center}
\includegraphics[scale=0.5]{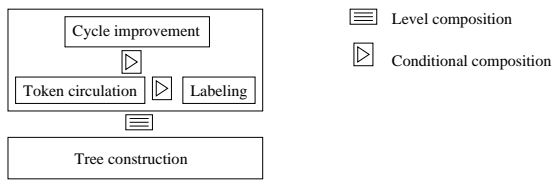}
\end{center}
\caption{Composition of the presented modules.}
\label{fig:composition}
\end{figure}

All the different modules presented, except the tree construction parts of the correction module, need the presence of a spanning tree in $G$. Thus, we must execute the tree construction rules first if an incoherency in the spanning tree is detected. To this end, these rules are composed using the level composition defined in \cite{GoudaH91}. If Predicate $\CoherentTree(v)$ is not verified then the tree construction rules are executed, otherwise the other modules can be executed.
The token circulation algorithm and the naming algorithm are composed together using the conditional composition described in \cite{DGPV01}. Finally, we compose the token circulation algorithm and the cycle improvement module with a conditional composition using Predicate $\CDFS(v)$ defined in the algorithm. This allows to execute the token circulation algorithm only if the cycle improvement module does not need the token on a node.
Figure \ref{fig:composition} shows how the different modules are composed together.

\section{Concluding remarks}

We presented a new solution to the distributed MST construction that is both self-stabilizing and loop-free. It improves on memory usage from $O(n \log n)$ to $O(\log n)$, yet doesn't make strong system assumptions such as knowledge of network size or unicity of edge weights, making it particularly suited to dynamic networks. Two important open questions are raised:
\begin{enumerate}
\item For depth first search tree construction, self-stabilizing solutions that use only constant memory space do exist. It is unclear how the obvious constant space lower bound can be raised with respect to metrics that minimize a global criterium (such as MST).
\item Our protocol pionneers the design of self-stabilizing loop-free protocols for \emph{non} locally optimizable tree metrics. We expect the techniques used in this paper to be useful to add loop-free property for other metrics that are only globally optimizable, yet designing a generic such approach is a difficult task.
\end{enumerate}

\newpage
\pagenumbering{roman}


\begin{thebibliography}{10}

\bibitem{AH93c}
Efthymios Anagnostou and Vassos Hadzilacos.
\newblock Tolerating transient and permanent failures (extended abstract).
\newblock In Andr{\'e} Schiper, editor, {\em WDAG}, volume 725 of {\em Lecture
  Notes in Computer Science}, pages 174--188. Springer, 1993.

\bibitem{BDH08c}
Michael Ben-Or, Danny Dolev, and Ezra~N. Hoch.
\newblock Fast self-stabilizing byzantine tolerant digital clock
  synchronization.
\newblock In Rida~A. Bazzi and Boaz Patt-Shamir, editors, {\em PODC}, pages
  385--394. ACM, 2008.

\bibitem{CG02j}
Jorge~Arturo Cobb and Mohamed~G. Gouda.
\newblock Stabilization of general loop-free routing.
\newblock {\em J. Parallel Distrib. Comput.}, 62(5):922--944, 2002.

\bibitem{DGPV01}
Ajoy~Kumar Datta, Shivashankar Gurumurthy, Franck Petit, and Vincent Villain.
\newblock Self-stabilizing network orientation algorithms in arbitrary rooted
  networks.
\newblock {\em Stud. Inform. Univ.}, 1(1):1--22, 2001.

\bibitem{D74j}
Edsger~W. Dijkstra.
\newblock Self-stabilizing systems in spite of distributed control.
\newblock {\em Commun. ACM}, 17(11):643--644, 1974.

\bibitem{D00b}
S.~Dolev.
\newblock {\em Self-stabilization}.
\newblock MIT Press, March 2000.

\bibitem{Dolev00}
Shlomi Dolev.
\newblock {\em Self-Stabilization}.
\newblock MIT Press, 2000.

\bibitem{DH97j}
Shlomi Dolev and Ted Herman.
\newblock Superstabilizing protocols for dynamic distributed systems.
\newblock {\em Chicago J. Theor. Comput. Sci.}, 1997, 1997.

\bibitem{DW97j}
Shlomi Dolev and Jennifer~L. Welch.
\newblock Wait-free clock synchronization.
\newblock {\em Algorithmica}, 18(4):486--511, 1997.

\bibitem{DW04j}
Shlomi Dolev and Jennifer~L. Welch.
\newblock Self-stabilizing clock synchronization in the presence of byzantine
  faults.
\newblock {\em J. ACM}, 51(5):780--799, 2004.

\bibitem{Gafni81}
Eli~M. Gafni and P.~Bertsekas.
\newblock Distributed algorithms for generating loop-free routes in networks
  with frequently changing topology.
\newblock {\em IEEE Transactions on Communications}, 29:11--18, 1981.

\bibitem{GallagerHS83}
Robert~G. Gallager, Pierre~A. Humblet, and Philip~M. Spira.
\newblock A distributed algorithm for minimum-weight spanning trees.
\newblock {\em ACM Trans. Program. Lang. Syst.}, 5(1):66--77, 1983.

\bibitem{Aceves93}
J.~J. Garcia-Luna-Aceves.
\newblock Loop-free routing using diffusing computations.
\newblock {\em IEEE/ACM Trans. Netw.}, 1(1):130--141, 1993.

\bibitem{GP93c}
Ajei~S. Gopal and Kenneth~J. Perry.
\newblock Unifying self-stabilization and fault-tolerance (preliminary
  version).
\newblock In {\em PODC}, pages 195--206, 1993.

\bibitem{GoudaH91}
Mohamed~G. Gouda and Ted Herman.
\newblock Adaptive programming.
\newblock {\em IEEE Trans. Software Eng.}, 17(9):911--921, 1991.

\bibitem{GS99c}
Mohamed~G. Gouda and Marco Schneider.
\newblock Stabilization of maximal metric trees.
\newblock In Anish Arora, editor, {\em WSS}, pages 10--17. IEEE Computer
  Society, 1999.

\bibitem{AntonoiuS97}
Sandeep K.~S. Gupta and Pradip~K. Srimani.
\newblock Self-stabilizing multicast protocols for ad hoc networks.
\newblock {\em J. Parallel Distrib. Comput.}, 63(1):87--96, 2003.

\bibitem{HighamL01}
Lisa Higham and Zhiying Liang.
\newblock Self-stabilizing minimum spanning tree construction on
  message-passing networks.
\newblock In {\em DISC}, pages 194--208, 2001.

\bibitem{JT03c}
Colette Johnen and S{\'e}bastien Tixeuil.
\newblock Route preserving stabilization.
\newblock In Shing-Tsaan Huang and Ted Herman, editors, {\em Self-Stabilizing
  Systems}, volume 2704 of {\em Lecture Notes in Computer Science}, pages
  184--198. Springer, 2003.

\bibitem{JT03}
Colette Johnen and S{\'e}bastien Tixeuil.
\newblock Route preserving stabilization.
\newblock In {\em Self-Stabilizing Systems}, pages 184--198, 2003.

\bibitem{Kruskal56}
Joseph~B. Kruskal.
\newblock On the shortest spanning subtree of a graph and the travelling
  salesman problem.
\newblock {\em Proc. Amer. Math. Soc.}, 7:48--50, 1956.

\bibitem{PT97j}
Marina Papatriantafilou and Philippas Tsigas.
\newblock On self-stabilizing wait-free clock synchronization.
\newblock {\em Parallel Processing Letters}, 7(3):321--328, 1997.

\bibitem{PetitV07}
Franck Petit and Vincent Villain.
\newblock Optimal snap-stabilizing depth-first token circulation in tree
  networks.
\newblock {\em J. Parallel Distrib. Comput.}, 67(1):1--12, 2007.

\bibitem{Prim57}
R.C. Prim.
\newblock Shortest connection networks and some generalizations.
\newblock {\em Bell System Tech. J.}, pages 1389--1401, 1957.

\bibitem{Tarjan83}
Daniel~Dominic Sleator and Robert~Endre Tarjan.
\newblock A data structure for dynamic trees.
\newblock {\em J. Comput. Syst. Sci.}, 26(3):362--391, 1983.

\end{thebibliography}

\newpage
\appendix
\section*{Appendix}

\subsection*{Correctness proof}



We use the algorithm given in \cite{JT03} to construct a breadth first search spanning tree. Note that, the algorithm 
given in \cite{JT03} satisfies the loop-free property. Therefore, in the remainder we suppose there 
is a constructed spanning tree.

\begin{theorem}[\LFMST]
Starting from an arbitrary spanning tree of the network $G$, $\LFMST$ algorithm 
is a self-stabilizing loop-free algorithm.
\end{theorem}

\begin{proof}
Let $T$ a spanning tree of network $G$ and $v$ a node of $T$. If $v$ is in an incoherent state then according to Lemma \ref{lem:correct_node} below, the algorithm bootstraps the state of $v$, otherwise the token continues its circulation in the tree until a verification on a node is needed (Lemma \ref{lem:token}). When the token is on a node that has 
candidate edges not in the tree (i.e. whose fundamental cycle is not yet checked), according to Corollary \ref{cor:verif_cycle_noeud} the algorithm verifies if an amelioration (see Section \ref{sec:high} for the definition of an amelioration) must be performed using these not tree edges and according to Lemma \ref{lem:do_improve} an improvement is performed if an improvement is possible. Moreover, the algorithm performs all possible improvements (Lemma \ref{lem:do_possible_improve}) until no improvement is feasable (Lemma \ref{lem:convergence} and Corollary \ref{cor:always_mst}), i.e. a minimum spanning tree is reached.

Starting from a spanning tree $T$ of the network, during the execution of the algorithm no cycle is created and a spanning tree structure is preserved (see Corollary \ref{cor:no_cycle}). Moreover, according to Lemma \ref{lem:correction} if $T$ is minimum spanning tree then $T$ is maintained by the algorithm.
\end{proof}



\begin{lemma}[Bootstrap]
\label{lem:correct_node}
A node $v$ in an incoherent state for the cycle improvement module eventually verifies the predicate $\CoherentCycle(v)$.
\end{lemma}

\begin{proof}
A node may have six different states in the algorithm: $\Do, \Ve, \Imp, \End, \Err$, and $\Prop$. The coherence of a node in these different states is defined respectively by predicates $\CoherentD, \CoherentV, \CoherentI, \CoherentEI$, and $\CoherentE$. For the state $\Prop$, we detect if the propagation is done using Predicate $\EndProp(v)$ to allow the execution of Rule $\RCA$ to reinitialize the state of the node. According to the algorithm description, if a node $v$ is not coherent (i.e. does not respect one of the previous mentioned predicates), Predicate $\CoherentCycle(v)$ is not verified since the previous mentioned predicates are exclusive because a node can have one state. Thus, $v$ can execute Rule $\RCA$ to correct its variables to a coherent state satisfying Predicate $\CoherentD(v)$. As a consequence Predicate $\CoherentCycle(v)$ is satisfied too (see Rule $\RCA$).
\end{proof}

\begin{lemma}
\label{lem:label_error}
If $\CoherentCycle(v)=true$, $\NIC(v)=\emptyset$ and $\EndProp(v)=true$ then eventually a node $v$ is in status $\Err$ and satisfies $\CoherentE(v)$.
\end{lemma}

\begin{proof}
We show that if a node $v$ in a fundamental cycle has no successor because of bad labels then $v$ changes its status to $\Err$. Predicate $\NIC(v)$ is in charge to give the successor of a node in a fundamental cycle based on the node labels, following Observation \ref{lem:cycle_struct} below. Thus, if Predicate $\NIC(v)$ returns no successor this implies that bad labels disturb the computation of the successor. Predicate $\Error(v)$ is in charge to detect bad labels. We show that a node $v$ which is part of a fundamental cycle (i.e. satisfies Predicate $\CoherentCycle(v)$) and detects an error or has its successor in status $\Err$ changes its status to $\Err$ (except the initiator node, i.e. $\DefC[0]_v \neq \La_v$). We do not consider the status $\Do$ since in this status no node has a successor (see Predicate $\Error(v)$.

Consider any node $v$ (except the initiator node) which satisfies Predicate $\CoherentCycle(v)$. To change its status to $\Err$ a node must execute Rule $\RCB$ and we must consider two cases: a node with no successor, or a node with a successor in status $\Err$. In the first case, a node $v$ satisfies Predicate $\Error(v)$ (see Predicate $\Error(v)$) and $v$ can execute Rule $\RCB$. After the execution of Rule $\RCB$, $v$ satisfies Predicate $\CoherentE(v)$ since $\state_v=\Err$, $\NIC(v)=\emptyset$ and $\DefC_v=\DefC_{\pred(v)}$. In the second case, suppose that for a node $v$ we have $\state_{\next_v}=\Err$. According to Predicate $\AskE(v)$, $\Error(v)=true$ and thus $v$ can execute Rule $\RCB$ to change its status to $\Err$. After the execution of Rule $\RCB$, $v$ satisfies Predicate $\CoherentE(v)$ since $\state_v=\Err$, $\AskE(v)=true$ and $\DefC_v=\DefC_{\pred(v)}$. One can show by induction following the same argument that any node part of a fundamental cycle with bad labels changes its status to $\Err$ (except the initiator node).
\end{proof}

\begin{lemma}[Token circulation]
\label{lem:token}
Starting from a configuration where a spanning tree $T$ is constructed, if a node $v$ has the DFS token and satisfies $\CoherentCycle(v)$ then eventually Predicate $\CDFS(v)$ returns true.
\end{lemma}

\begin{proof}
Predicate $\CDFS(v)$ notices when the DFS token must continue its circulation in the tree. The DFS token must continue its circulation in four cases: (1) a node in status $\Do$ has no candidate edge, (2) a node in status $\Ve$ with no possible improvement has no candidate edge, (3) an improvement was done in the fundamental cycle, or (4) bad labels are detected in the fundamental cycle.

In case 1, for node $v$, $\CoherentD(v)=true$ (otherwise according to Lemma \ref{lem:correct_node} its state is reinitialized). In case 2, for node $v$, $\CoherentV(v)=true$ (otherwise according to Lemma \ref{lem:correct_node} its state is reinitialized) and Predicate $\ImproveF(v)$ is used to detect possible improvements (see the proof of Lemma \ref{lem:verif_cycle}). For case 1 and 2, if $v$ has no candidate Predicate $\Candidate(v)=false$ (see Predicate $\Candidate(v)$ and proof of Lemma \ref{lem:verif_cycle}) and thus Predicate $\CDFS(v)$ is satisfied. In case 3, according to Lemma \ref{lem:end_improve} the initiator node $v$ satisfies Predicate $\CoherentEI(v)$ and Predicate $\CDFS(v)$ returns true. Finally in case 4, according to Lemma \ref{lem:label_error} the successor of the initiator node $v$ is in status $\Err$ so Predicate $\AskE(v)=true$ and Predicate $\Error(v)$ returns true. Thus, Predicate $\CDFS(v)$ returns true.

Therefore, in all the above cases Predicate $\CDFS(v)$ returns true and $v$ can execute Rule $\RDFS$ to allow the token circulation. It then changes its status to $\Do$ and sets $\DefC[1]_v$ to $\done$ to force the verification of all adjacent candidate edges in the next tree traversal by the DFS token.
\end{proof}

\begin{observation}
\label{lem:cycle_struct}
Let $T$ be a tree spanning $V$ and correctly labeled. Let an edge $e=\{u,v\} \in E, e \not \in T$, $C_e$ its fundamental cycle and $x$ the fundamental cycle root of $C_e$. There is always a path $P(u,v)$ in $T$ between $u$ and $v$, such that $P(u,v)$ can be decomposed in two parts: a sub-path $P(x,u) \subset P(u,v)$ (resp. $P(x,v) \subset P(u,v)$) with increasing labels from $x$ to $u$ (resp. $x$ to $v$).
\end{observation}

\begin{lemma}[Cycle verification]
\label{lem:verif_cycle}
Let $v$ be a node of $T$ such that $v$ has the DFS token with at least an adjacent edge $e=\{u,v\} \in E, e \not \in T$ whose fundamental cycle is not already verified by the algorithm. Eventually the cycle improvement module verifies if there is an improvement in $C_e$.
\end{lemma}

%

\begin{proof}
Suppose first that $v$ has the DFS token and $v$ is in a coherent state $\Do$, otherwise according to Lemma \ref{lem:correct_node} its state is corrected. Let $e=\{u,v\}$ be a not tree edge, which is a candidate edge for $v$, i.e., we have $\Candidate(v) \neq end$. We consider that $\La_u < \La_v$ since a candidate edge for node $v$ is an adjacent not tree edge $e=\{u,v\}$ with $\La_u < \La_v$, see predicate $\LCandidate(v)$. Since $v$ is in a coherent state $\Do$ and $\Candidate(v) \neq end$, we have variable $\DefC[1]_v$ equal to $\done$, Predicate $\CoherentCycle(v)$ and $\InitV(v)$ return true, whears Predicate $\Error(v)$ returns false. Thus, $v$ can execute Rule $\RV$. Note that Rule $\RDFS$ can not be executed since Predicate $\CDFS(v)$ returns false since $\Candidate(v) \neq end$. As a consequence $v$ stops the DFS token and becomes the initiator node of cycle $C_e$ with $u$ as target node (see Rule $\RV$).

After the execution of Rule $\RV$, $v$ is in state $\Ve$ and according to predicate $\NIC(v)$ $v$ selects its father as next node in the cycle (i.e. $\next_v=\parent_v$). Note that since $v$ is in coherent state $\Do$ variable $\VarC_v=(0,\Be)$. Cycle $C_e$ is decomposed in two parts (see Lemma \ref{lem:cycle_struct}): (1) from the initiator $v$ to the root $x$ of $C_e$ and (2) from $x$ to the target node $u$. In the following we prove by induction on the length of cycle $C_e$ that a node $a$ belonging to $C_e$ executes Rule $\RV$ and eventually is in state $\Ve$. Moreover, variable $\next_a$ describes the successor of $a$ in $C_e$ (i.e. encodes the cycle $C_e$).

Case 1: Consider a coherent node $a$ in state $\Do$ (see Lemma \ref{lem:correct_node}) which has not the DFS token (i.e. Predicate $\Init(a)$ is false). Consider the successor node of $C_e$'s initiator node $v$. As described above, $v$ is in state $\Ve$ and $\next_v=a$. According to Predicate $\pred(a)$, $v$ is the predecessor of $a$ in cycle $C_e$ since $a$ is the parent of $v$ in the tree. Thus, Predicate $\AskV(a)$ returns true and $a$ could execute Rule $\RV$. Therefore, $a$ is in state $\Ve$ and selects its parent as its successor in $C_e$, like $v$. Moreover, $a$ computes the new heaviest edge from $v$ to $a$ and notices that the heaviest edge location is before (i.e. $\Be$) the root of $C_e$ (see respectively predicates $\MaxC$ and $\WayC$). Using the same scheme, we can show that all nodes on $C_e$ between $v$ and $x$ (including $x$) execute Rule $\RV$ and are in state $\Ve$.

Case 2: Consider a coherent node $a$ in state $\Do$ (see Lemma \ref{lem:correct_node}) which has not the DFS token (i.e. Predicate $\Init(a)$ is false) and is the successor node of $x$. As described in case 1, $x$ is in state $\Ve$. Since $x$ is the parent of $a$ in the tree, Predicate $\pred(a)$ returns $x$ as predecessor of $a$. Thus, Predicate $\AskV(a)$ returns true and $a$ can execute Rule $\RV$. $a$ selects as its successor in $C_e$ the child with the highest label smaller than target node's $u$ label (see predicates $\maxLab(a)$ and $\NIC(a)$). Moreover, $a$ computes the new heaviest edge from $v$ to $a$ and if $a$ has a different heaviest edge $a$ notice that the heaviest edge location is after (i.e. $\Af$) the root of $C_e$ otherwise $a$ takes the location of its predecessor (see respectively predicates $\MaxC$ and $\WayC$). Using the same scheme, we can show that all nodes on $C_e$ between $x$ and $u$ (including $u$) execute Rule $\RV$ and are in state $\Ve$. Note that the target node $u$ selects $v$ as its successor in $C_e$ (see Predicate $\NIC(u)$).

Consider now that $v$ has the DFS token, is in a coherent state $\Ve$ and predecessor of $v$ is in state $\Ve$ (i.e. $\AskV(v)=true$). Note that the predecessor of $v$ is the target node $u$. As described in case 2, target node $u$ knows the weight of the heaviest edge $e'$ in $C_e$ ($e' \in T$). Thus, $v$ could check if there is an improvement in $C_e$ (see Predicate $\Improve(v)$). 
\end{proof}

\begin{corollary}[Node cycles verification]
\label{cor:verif_cycle_noeud}
Let $T$ a spanning tree and $v$ be a node of $T$ such that $v$ has the DFS token. Eventually for each adjacent candidate edge $e$ of $v$, the cycle improvement module verifies if there is an improvement in $C_e$.
\end{corollary}

\begin{proof}
We prove that while there is no improvement initiated by $v$, each edge $e=\{u,v\} \in E, e \not \in T$ is eventually examined by the cycle improvement module. We 
consider the two cases below: (1) there is no improvement initiated by $v$, or (2) an improvement can be done in $C_e$ for a candidate edge $e$. Consider an arbitrary candidate edge $e=\{u,v\} \in E, e \not \in T$. According to Lemma \ref{lem:verif_cycle}, $v$ eventually verifies if there is an improvement in $C_e$.

Case 1: If there is no improvement in $C_e$ and $v$ has another candidate edge (i.e. predicates $\Candidate(v)$ and $\InitV(v)$ return true) then $v$ must check if there is an improvement in the fundamental cycle of the new candidate edge. According to Lemma \ref{lem:verif_cycle}, $v$ could execute again Rule $\RV$ with a new target and stay in a coherent state $\Ve$. Therefore for each not tree adjacent edge $e$, $v$ eventually verifies if there is an improvement in the fundamental cycle $C_e$.

Case 2: If an improvement can be done in $C_e$, when the improvement is done, $v$ is in the state $\EndI$. Thus, Predicate $\CDFS(v)$ returns true and Rule $\RDFS$ can be executed to continue the token circulation in the tree. However, the next time $v$ has the token as described in case 1, $v$ eventually checks again the previously examined edges, 
but $v$ will also check candidate edges not previously visited.

\end{proof}


\begin{definition}[Red Rule]
\label{def:red_rule}
If $C$ is a cycle in $G=(V,E)$ with no red edges then color in red the maximum weight edge in $C$.
\end{definition}

\begin{theorem}[Tarjan et al. \cite{Tarjan83}]
\label{thm:tarjan_mst}
Let $G$ be a connected graph. If it is not possible to apply Red Rule then the set of not colored edges forms a minimum spanning tree of $G$.
\end{theorem}

\begin{lemma}[Improvement]
\label{lem:do_improve}
Let an edge $e=\{u,v\} \in E, e \not \in T$ and let $C_e$ its fundamental cycle. If there exists a possible improvement in 
$C_e$ then the algorithm eventually performs the improvement.
\end{lemma}

\begin{proof}
According to Definition \ref{def:red_rule}, there is an improvement in a cycle $C$ if the edge of maximum weight in $C$ belongs to the current tree and one can use the Red Rule. Given an edge $e=\{u,v\} \in E, e \not \in T$ and $C_e$ its fundamental cycle, Lemma \ref{lem:verif_cycle} states that the initiator node $v$ detects if there is an improvement in cycle $C_e$. Assume that an improvement can be performed in cycle $C_e$ (i.e. predicate $\Improve(v)=true$). As proved in Lemma \ref{lem:verif_cycle}, $u$ and $v$ are in a coherent state $\Ve$ and have a successor, thus we have $\CoherentCycle(v)=true, \Error(v)=false$ and $\AskV(v)=true$. Since $v$ is the initiator node of $C_e$, $v$ has the DFS token and could not be the root of $C_e$ (i.e. $\Forward(v)=true$ and $\CA(v)=false$). So $v$ can execute Rule $\RIB$, to change its state to $\Imp$ and to update its estimation of the heaviest edge of $C_e$ and the heaviest edge location to the values of its predecessor (i.e. the target node $u$). Two cases have to be analyzed: (1) the heaviest edge location is between $v$ and $x$ (i.e. $\VarC[1]_v=\Be$) or (2) between $u$ and $x$ (i.e. $\VarC[1]_v=\Af$). In the two cases, the improvement must be propagated from $v$ to $x$ (resp. $u$ to $x$) until reaching the (first) heaviest edge or the root of $C_e$ (if the weight of the heaviest edge has been reduced). Indeed, the root of $C_e$ must not change its parent to a neighbor in $C_e$ otherwise it disconnects its subtree from the rest of the tree.

Case 1: Since $\VarC[1]_v=\Be$, $v$ takes as new parent its predecessor in the cycle. Let $a$ be a node in coherent state $\Ve$ between $v$ and $x$ (Note: $a$ exists otherwise suppose $a$ is in an incoherent state, according to Lemma \ref{lem:correct_node} $a$ reinitiates its state to $\Do$ which induces a propagation of state $\Do$ in $C_e$, since the nodes are no more coherent with their predecessors, and stops the improvement until a new verification of $C_e$ is restarted). If the improvement must continue (i.e. Predicate $\Improve(a)$ returns true), $a$ is not the root of $C_e$ and its predecessor is in state $\Imp$ (see Predicate $\AskI$) then $a$ can execute Rule $\RIB$. So, $a$ changes its state to $\Imp$, updates its variable $\VarC_a$ to the value of its predecessor and takes its predecessor as its parent. This propagation continues until reaching a node $a$ which stops the improvement (i.e. $\Improve(a)=false$ or $\CA(a)=true$).

Case 2: $\VarC[1]_v=\Af$ and as in case 1 $v$ executes Rule $\RIB$ but $v$ changes only its state to $\Imp$ and updates its variable $\VarC_v$ to the value of its predecessor. Hence $v$ does not change its parent. Consider the target node $u$, we have $\AskI(u)=true$ since $v$ is in state $\Imp$. So, $u$ executes Rule $\RIB$, changes its state to $\Imp$, updates $\VarC_u$ to its successor value and changes its parent to its successor (i.e. $\parent_u=v$). As described in case 1, the improvement is propagated in the cycle from $u$ to $x$ until a node $a$ is reached which stops the improvement (i.e. $\Improve(a)=false$ or $\CA(a)=true$).

Overall, if an improvement exists then this improvement is eventually performed.
\end{proof}

\begin{lemma}
\label{lem:end_improve}
If $v$ satisfies $\CoherentI(v)$ and $\EndProp(v)$ then $v$ eventually changes its status to $\EndI$ and the predicate $\CoherentEI(v)$ is satisfied.
\end{lemma}

\begin{proof}
We conduct the proof by induction on the length of the fundamental cycle. A node involved in an improvement executes Rule $\RIE$ to inform its predecessor or successor the end of the improvement. An improvement can be propagated by a successor or a predecessor in the cycle. We show the lemma considering that the improvement is propagated by the successor of a node, but the same idea can be applied by considering predecessor instead of successor. Moreover, we assume that labels are correct in the fundamental cycle otherwise it is not necessary to inform the end of the improvement since according to Lemma \ref{lem:label_error} the nodes are in state $\Err$. Let $x$ the node which detects the end of the improvement and $y$ the initiator node in a fundamental cycle.

Consider the node $x$, such that $\CoherentI(x)=true$ and $w(x,\next_x) \geq \VarC[0]_x$. Predicate $\EndImprove(x)=true$ since $\CoherentI(x)=true$ and $\NdDelete(x)$ is satisfied because $\Improve(x)=false$. Thus, $x$ can execute Rule $\RIE$ and changes its status to $\EndI$. Therefore, $\CoherentEI(x)$ is satisfied since $\state_x=\EndI$, $\NdDelete(x)=true$ and $\DefC_x=\DefC_{\parent_x}$ because $x$ and its parent are in the same fundamental cycle. Now, suppose by induction hypothesis that any node $u$ between $x$ and the initiator node $y$ are in state $\EndI$ and $\CoherentEI(u)$ is satisfied. Consider the initiator node $y$, $\state_y=\Imp$, $\CoherentI(y)=true$ and $\state_{\next_y}=\EndI$. Predicate $\EndImprove(y)$ is satisfied because Predicate $\AskEI(y)=true$ since $\state_{\next_y}=\EndI$ and $\DefC_y=\DefC_{\next_y}$. Thus, $y$ can execute Rule $\RIE$ and changes its status to $\EndI$. Therefore, Predicate $\CoherentEI(y)$ is satisfied since $\state_y=\EndI$, $\AskEI(y)=true$ and $\DefC_y=\DefC_{\parent_y}$ because $y$ and its parent are in the same fundamental cycle.
\end{proof}

%

\begin{lemma}[MST construction]
\label{lem:do_possible_improve}
Given a spanning tree $T$, the cycle improvement module performs an improvement if $T$ is not a minimum spanning tree of $G$.
\end{lemma}

\begin{proof}
According to the token circulation algorithm \cite{PetitV07}, eventually each node in the tree is visited and holds the token. 
Consider a node $v$ in the tree $T$, which has the DFS token. According to Corollary \ref{cor:verif_cycle_noeud} 
eventually each adjacent candidate edge of $v$ is examined by the cycle improvement module. Thus, if an improvement is possible this one is detected according to Lemma \ref{lem:verif_cycle} and performed by $v$ according to Lemma \ref{lem:do_improve}. Therefore, if an improvement is possible the cycle improvement module performs it.
\end{proof}

\begin{lemma}
\label{lem:no_improve}
Let $T$ be an existing minimum spanning tree of $G$. The algorithm performs no improvement.
\end{lemma}

\begin{proof}
Let $T$ be an existing minimum spanning tree of $G$ and $v$ be a node in $T$ which has the DFS token. Let $e=\{u,v\}, e \not \in T$ an adjacent candidate edge of $v$ and $C_e$ its corresponding fundamental cycle. Suppose the cycle improvement module performs an improvement in $C_e$. We prove by contradiction that no improvement could performed by the algorithm.

Let $w(C_e)$ the maximum edge weight in $C_e$, excluding edge $e$. According to Definition \ref{def:red_rule}, to initiate an improvement from $v$ the following condition must be verified: $w(C_e)>w(e)$. According to Lemma \ref{lem:verif_cycle}, the predecessor $u$ of $v$ holds the maximum edge weight in $C_e$ (i.e. $\VarC[0]_u=w(C_e)$). To perform an improvement, Predicate $\Improve(v)$ must return true to allow $v$ to execute Rule $\RIB$. This implies that $\max(\VarC[0]_v,\VarC[0]_u)>w(u,v)$ (see Predicate $\Improve(v)$), i.e. $w(C_e)>w(u,v)$ (since $\VarC[0]_u=w(C_e)$) which contradicts the fact that no improvement can be performed in $C_e$. Therefore, $v$ can not execute Rule $\RIB$ if no improvement is possible in a fundamental cycle.
\end{proof}

\begin{corollary}[MST conservation]
\label{cor:always_mst}
Let $T$ be an existing minimum spanning tree of $G$. The algorithm maintains a spanning tree.
\end{corollary}

\begin{proof}
Lemma \ref{lem:no_improve} shows that no improvement is performed by the algorithm if $T$ is a minimum spanning tree of $G$, i.e. Rule $\RIB$ can not be executed by a node. Therefore, according to Lemma \ref{lem:no_improve} and by Remark \ref{rem:parent_change} a spanning tree is maintained.
\end{proof}

\begin{lemma}[Convergence]
\label{lem:convergence}
Starting from an illegitimate configuration eventually the algorithm reaches in a finite time a legitimate configuration.
\end{lemma}

\begin{proof}
If the initial configuration contains no spanning tree, there is a node $v$ such that Predicate $\CoherentTree(v)=false$ and according to the level composition (defined in \cite{GoudaH91}) we use the algorithm given in \cite{JT03} to construct a breadth first search spanning tree. Otherwise, the initial configuration contains a spanning tree which is not a minimum spanning tree. According to Lemma \ref{lem:do_possible_improve} and \ref{lem:no_improve}, improvements are performed by the cycle improvement module until a minimum spanning tree is reached. Moreover, according to Lemma \ref{lem:no_disconnect_tree} a spanning tree is preserved by the cycle improvement module. Finally, there is at most $m-n+1$ fundamental cycles in any graph so a finite number of improvements can be performed by the cycle improvement module. Thus, in a finite time the algorithm returns a minimum spanning tree.
\end{proof}

\begin{remark}
\label{rem:parent_change}
According to the cycle improvement module description, only Rule $\RIB$ could change the parent of a node.
\end{remark}

\begin{lemma}
\label{lem:no_disconnect_tree}
Let $T$ be an existing tree spanning $V$, no move performed by the cycle improvement module disconnects $T$.
\end{lemma}

\begin{proof}
There is two cases in which the existing tree $T$ spanning $V$ is disconnected. It is necessary (1) to delete an edge of $T$ by changing the parent of a node (except the root of $T$) to itself or (2) to attribute as new parent of a node a neighbor belonging to its descendant in the tree. Consider the execution of Rule $\RIB$ (see Remark \ref{rem:parent_change}). Rule $\RIB$ can be executed by a node if this one is in state $\Ve$ and is a coherent node (see predicate $\CoherentV$ in Rule $\RIB$). As described in the proof of Lemma \ref{lem:verif_cycle}, a coherent node in state $\Ve$ has a predecessor and a successor in a fundamental cycle, note that the initiator has a predecessor because it must wait that this one (i.e. the target node) is in state $\Ve$ to execute Rule $\RIB$ (see predicate $\AskV$).


Case (1) is not permitted by the algorithm. The new parent of a node is its predecessor or successor in the fundamental cycle (see Rule $\RIB$). Thus the algorithm selects as new parent another node different of the node itself.

Case (2) is not permitted by the algorithm, since the new parent of a node executing Rule $\RIB$ is its predecessor or successor in the fundamental cycle and the edge between the node and its new parent is not already in the tree (see predicate $\Improve$). In other words, the algorithm adds and deletes two adjacent edges in the fundamental cycle, which gives after each move a new spanning tree. 
Moreover, the algorithm can not change the parent of a fundamental cycle root (see predicate $\CA$ in guard of Rule $\RIB$), in particular the root of the tree, otherwise the subtree of the fundamental cycle root could be disconnected from the rest of the tree. Thus, the new parent is an ancestor or another node with the same ancestor in the tree.

Therefore, after each move performed by the algorithm a spanning tree is preserved.
\end{proof}

\begin{corollary}[Loop-free property]
\label{cor:no_cycle}
Let $T$ be an existing tree spanning $V$, after any move performed by the cycle improvement module $Cycle(T,u,v)=false, \forall u,v \in V$.
\end{corollary}

\begin{proof}
In a configuration where a spanning tree $T$ is constructed, we have $Cycle(T,u,v)=false, \forall u,v \in V$ otherwise it contradicts the fact that $T$ is a spanning tree. Moreover, according to Case (2) in the proof of Lemma \ref{lem:no_disconnect_tree} any move of the cycle improvement module preserves a spanning tree structure. Thus, for any move $Cycle(T,u,v)=false, \forall u,v \in V$.
\end{proof}

\begin{lemma}[Closure]
\label{lem:correction}
Starting from a legitimate configuration the algorithm preserves a legitimate configuration.
\end{lemma}

\begin{proof}
Let $T$ be an existing tree spanning $V$, such that $T$ is a minimum spanning tree of $G$. Thus, $\forall v \in V, \CoherentTree(v)=true$. According to the level composition (defined in \cite{GoudaH91}), since on a node $v$ the predicate $\CoherentTree(v)$ determines if the tree must be reconstructed, the only modules executed are the token circulation with labeling module given respectively in \cite{PetitV07,DGPV01} and the cycle improvement module. The conditional composition (defined in \cite{DGPV01}) between the token circulation with labeling module and the cycle improvement module, using Predicate $\CDFS(v)$ on a node $v$ determines which module has to be executed. According to Lemma \ref{lem:token}, for any node $v \in V$ eventually Predicate $\CDFS(v)=true$ and the DFS token continue its circulation. Otherwise, only the cycle improvement module is executed. According to Lemma \ref{lem:no_improve} and Corollary \ref{cor:always_mst}, a minimum spanning tree of $G$ is preserved by the cycle improvement module and therefore by the algorithm composed of the different modules.
\end{proof}

\subsection*{Complexity}

\begin{lemma}
\label{lem:complex_gen}
Starting from a configuration where an arbitrary spanning tree is constructed, in at most $O(mn)$ rounds the cycle improvement module produces a minimum spanning tree of $G$, with respectively $m$ and $n$ the number of edges and nodes of the network $G$.
\end{lemma}

\begin{figure}
\begin{center}
\includegraphics[scale=0.8]{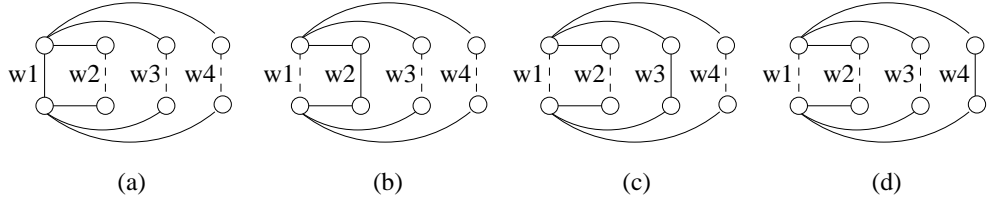}
\end{center}
\caption{(a) a spanning tree with plain lines in a graph with $m-n+1$ improvements, (b) the spanning tree obtained after the first improvement, (c) the spanning tree obtained after the second improvement, (d) the minimum spanning tree of the graph obtained after the third improvement.}
\label{fig:ex_complexite}
\end{figure}

\begin{proof}
In a given network $G=(V,E)$, if a spanning tree of $G$ is constructed then there are exactly $m-(n-1)$ fundamental cycles in $G$ since there are $n-1$ edges in any spanning tree of $G$. Thus, a tree edge can be contained in at most $m-n+1$ fundamental cycles. Consider a configuration where a spanning tree $T$ of $G$ is constructed and a tree edge $e_0$ is contained in $m-n+1$ fundamental cycles and all tree edges have a weight equal to 1, except $e_0$ of weight $w(e_0)>1$. Suppose that $T$ is not a minimum spanning tree of $G$ such that $\forall e_i \in E, i=1, \dots, m-n+1, w(e_{i-1})>w(e_i)$ with $e_0 \in T$ and $\forall i=1, \dots,m-n+1, e_i \not \in T$ and $w(e_i)>1$ (see the graph of Figure \ref{fig:ex_complexite}(a)). Consider the following sequence of improvements: $\forall i, i=1, \dots, m-n+1$, exchange the tree edge $e_{i-1}$ by the not tree edge $e_i$ (see a sequence of improvements in Figure \ref{fig:ex_complexite}). In this sequence, we have exactly $m-n+1$ improvements and this is the maximum number of improvements to obtain a minimum spanning tree since there are $m-n+1$ fundamental cycles and for each one we apply the Red rule (see Definition \ref{def:red_rule} and Theorem \ref{thm:tarjan_mst}). An improvement can be initiated in the cycle improvement module by a node with the DFS token. The DFS token performs a tree traversal in $O(n)$ rounds. Moreover, each improvement needs to cross a cycle a constant number of times and each cross requires $O(n)$ rounds. Since at most $m-n+1$ improvements are needed to obtain a minimum spanning tree, at most $O(mn)$ rounds are needed to construct a minimum spanning tree.
\end{proof}

\begin{lemma}
\label{lem:complex_change}
Starting from a legitimate configuration, after a weight edge modification the system reaches a legitimate configuration in at most $O(mn)$ rounds.
\end{lemma}

\begin{proof}
After a weight edge change the system is no more in a legitimate configuration in the following cases: (1) the weight of a not tree edge is less than the
 weight of the heaviest tree edge in its fundamental cycle, or (2) the weight of a tree edge is greater 
than the weight of a not tree edge in one of the fundamental cycles including the tree edges.

In each case above, the algorithm must verify if improvements must be performed to reach again a legitimate configuration, otherwise the system is still in a legitimate configuration. Thus, in case (1) it is only sufficient to verify if an improvement must be performed in the fundamental cycle associated to the not tree edge (i.e. to apply the Red rule a single time). To this end, its fundamental cycle must be crossed at most three times: the first time to verify if an improvement is possible, a second time to perform the improvement and a last time to end the improvement, each one needs at most $O(n)$ rounds. According to Lemma \ref{lem:do_improve} and \ref{lem:end_improve}, the improvement is performed by the algorithm which leads to a legitimate configuration. Case (2) is more complicated, indeed the weight of a tree edge can change which leads to a configuration where at most $m-n+1$ improvements must be performed to reach a legitimate configuration, since a tree edge can be contained in at most $m-n+1$ fundamental cycles as described in proof of Lemma \ref{lem:complex_gen}. Since each improvement phase needs $O(n)$ rounds (see case (1)) at most $O(mn)$ rounds are needed to reach a legitimate configuration.

The complexity of case (2) dominates the complexity of the first case. Therefore, after a weight edge change at most $O(mn)$ rounds are needed to reach a legitimate configuration.
\end{proof}

\end{document}